\documentclass[superscriptaddress,amsmath,amssymb,longbibliography]{revtex4-1}

\usepackage{graphicx}% Include figure files
\usepackage{dcolumn}% Align table columns on decimal point
\usepackage{bm}% bold math
\usepackage{amsmath}
\usepackage{tikz}
\usepackage{csquotes}
\usetikzlibrary{calc,shapes,arrows}
\newcommand {\Rey}{\mbox{\textrm{Re}}}
\begin{document}

%\preprint{AIP/123-QED}

\title[Couette-Poiseuille flow experiment with zero mean advection velocity: Subcritical transition to turbulence]{Couette-Poiseuille flow experiment with zero mean advection velocity: Subcritical transition to turbulence}% Force line breaks with \\
%\thanks{Footnote to title of article.}

\author{L. Klotz}
\email{lukasz.klotz@espci.fr}
\affiliation{Physique et M\'ecanique des Milieux H\'et\'erog\`enes (PMMH), CNRS - ESPCI - PSL Research University, 10 rue Vauquelin, 75005 Paris, France; Sorbonne Universit\'e, Universit\'e Paris Diderot}%

\author{G. Lemoult}
\affiliation{Institute of Science and Technology, Am Campus 1, 3400 Klosterneuburg, Austria}%

\author{I. Frontczak}%
\affiliation{Physique et M\'ecanique des Milieux H\'et\'erog\`enes (PMMH), CNRS - ESPCI - PSL Research University, 10 rue Vauquelin, 75005 Paris, France; Sorbonne Universit\'e, Universit\'e Paris Diderot}%
\affiliation{Institute of Aeronautics and Applied Mechanics, Warsaw University of Technology, Nowowiejska 24, 00-665 Warsaw, Poland}

\author{L.S. Tuckerman}
\affiliation{Physique et M\'ecanique des Milieux H\'et\'erog\`enes (PMMH), CNRS - ESPCI - PSL Research University, 10 rue Vauquelin, 75005 Paris, France; Sorbonne Universit\'e, Universit\'e Paris Diderot}%

\author{J.E. Wesfreid}
\email{wesfreid@pmmh.espci.fr}
\affiliation{Physique et M\'ecanique des Milieux H\'et\'erog\`enes (PMMH), CNRS - ESPCI - PSL Research University, 10 rue Vauquelin, 75005 Paris, France; Sorbonne Universit\'e, Universit\'e Paris Diderot}%

\date{\today}% It is always \today, today,
             %  but any date may be explicitly specified

\begin{abstract}
We present a new experimental set-up that creates a shear flow with zero mean advection velocity achieved by counterbalancing the nonzero streamwise pressure gradient by moving boundaries, which generates plane Couette-Poiseuille flow. We carry out the first experimental results in the transitional regime for this flow. Using flow visualization we characterize the subcritical transition to turbulence in Couette-Poiseuille flow and show the existence of turbulent spots generated by a permanent perturbation. Due to the zero mean advection velocity of the base profile, these turbulent structures are nearly stationary. We distinguish two regions of the turbulent spot: the active, turbulent core, which is characterized by waviness of the streaks similar to traveling waves, and the surrounding region, which includes in addition the weak undisturbed streaks and oblique waves at the laminar-turbulent interface. We also study the dependence of the size of these two regions on Reynolds number. Finally, we show that the traveling waves move in the downstream (Poiseuille) 
direction.
\end{abstract}

% PACS, the Physics and Astronomy % dodawac to?
%                             % Classification Scheme.
%                             % Valid PACS numbers may be entered using the \verb+\pacs{#1}+ command.
%\keywords{shear flows, subcritical transition to turbulence, plane Couette-Poiseuille flow}%Use showkeys class option if keyword
%        
                    %display desired
\maketitle
\section{Introduction}
\label{intro}
The transition to turbulence in wall bounded shear flows is characterized by the presence of localized turbulent regions containing coherent structures in the form of streamwise streaks \citep{schmid_stability_2001,landahl_a_1980}. These have been observed in different classical confined shear flows such as boundary layers \citep{gaster_experimental_1975,cantwell_structure_1978}, water tables \citep{emmons_laminar-turbulent_1951}, and pipe \citep{hof_turbulence_2005,mullin_experimental_2011}, channel \citep{lemoult_turbulent_2013} and plane Couette flows \citep{tillmark_experiments_1992,bottin_discontinuous_1998}. These turbulent structures are advected downstream with a speed that is approximately proportional to the bulk velocity. When this bulk velocity is non-zero, a very long test section is required to retain the turbulent spots for an appreciable time interval. Another difficulty is that these turbulent structures must be tracked as they move downstream.%tillmark_experimental_1991,

\begin{figure}
\begin{center}
\includegraphics[scale=0.9]{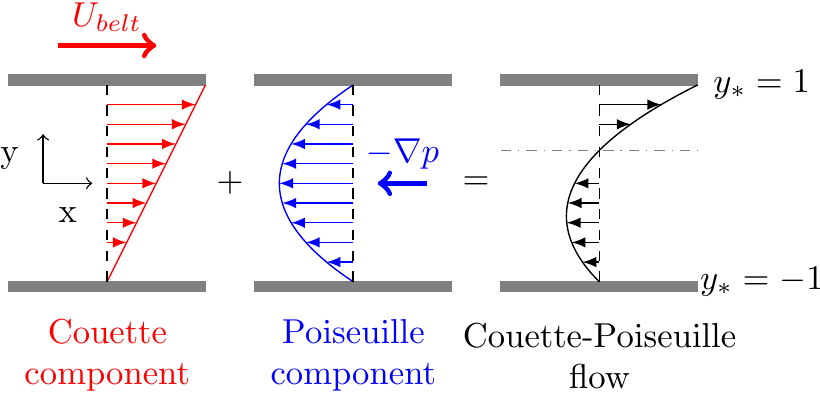}\\
\caption{Schematic representation of plane Couette-Poiseuille flow, which is a combination of Couette (in red) and Poiseuille (in blue) flows. The Couette flow is forced by the upper moving wall, whereas the Poiseuille flow is induced by streamwise pressure gradient. Dash-dotted gray line in the right subfigure separates the two inner regions of the flow: the upper one with $U>0$ dominated by the Couette component (high-shear region) and the lower one with $U<0$ dominated by the Poiseuille component (low-shear region).}
\label{fig:Scheme1}
\end{center}
\end{figure}

However, it is possible to cancel the mean flow velocity, as has been realized by pioneering experiments in plane Couette flow \citep{tillmark_experimental_1991,daviaud_subcritical_1992}. In these experimental set-ups, the base flow is induced by imposing opposite velocities at each wall of the test section, which generates a linear profile with zero mean velocity. If a turbulent spot is generated under such conditions, it remains stationary in the laboratory framework and there is no time limit on the observation of its evolution. The great advantage of a zero mean velocity has motivated us to construct a facility which is a generalization of the plane Couette experimental set-up. We combine the effect of one moving wall (which introduces a Couette component) and a streamwise pressure gradient due to the backflow generated by imposing zero mean flux rate (responsible for a Poiseuille component). The resulting base flow is a plane Couette-Poiseuille flow with zero mean advection velocity, shown in Fig.~\ref{fig:Scheme1}. To our knowledge, this is the first experimental investigation of subcritical transition to turbulence in plane Couette-Poiseuille flow.%

There are a number of theoretical results concerning Couette-Poiseuille flow. The linear stability analysis of this flow (necessarily two dimensional in the streamwise-cross-channel directions, due to Squire's theorem) was carried out \citep{potter_stability_1966, reynolds_finite-amplitude_1967, cowley_stability_1985, hains_stability_1967, drazin_hydrodynamic_1981, balakumar_finite-amplitude_1997, ozgen_heat_2006, savenkov_features_2010}, showing that when the Couette component is increased, the linear instability threshold shifts to higher values and the critical wave number decreases with respect to that of pure plane Poiseuille flow (see Fig.~4 in Ref. \onlinecite{balakumar_finite-amplitude_1997}). Even a relatively small component of Couette flow is sufficient to completely stabilize plane Poiseuille flow \citep{drazin_hydrodynamic_1981}. In this case, the linear instability threshold is infinite as it is for pure plane Couette flow. Specifically, it has been proved in Ref. \onlinecite{potter_stability_1966} that when the velocity of the Couette component exceeds $70\%$ of the center velocity of the Poiseuille component, the flow becomes stable to infinitesimal perturbations for all finite values of Reynolds number. This is in agreement with other results \citep{reynolds_finite-amplitude_1967, cowley_stability_1985} (note however, a slight difference of the coefficient values characterising the contribution of Couette and Poiseuille components reported in Ref. \onlinecite{drazin_hydrodynamic_1981}). Weakly nonlinear stability analysis was used to prove that, while it is stable to infinitesimal disturbances, Couette-Poiseuille flow is unstable to finite amplitude perturbations \citep{reynolds_finite-amplitude_1967,cowley_stability_1985,balakumar_finite-amplitude_1997,zhuk_asymptotic_2006}. The only fully nonlinear (but still two-dimensional) numerical study of transition to turbulence \citep{ehrenstein_two-dimensional_2008} used Poiseuille-Couette homotopy to continue a streamwise-localized finite-amplitude solution from Poiseuille to Couette flow. Another two dimensional study used a weakly nonlinear approach to investigate the time evolution of localized solutions in mariginally stable Couette-Poiseuille flow in the framework of the Ginzburg-Landau equation \citep{jennings_when_1999}. 

However, it is now believed that two dimensional evolution is not dynamically relevant for subcritical transition to three dimensional turbulence in shear flows. One of the features of subcritical transition to turbulence and three dimensional flow with streamwise or quasi-streamwise elongated streaks is transient linear growth. Its origin is the nonorthogonality of the linearized Navier-Stokes operator and the fact that streaks are the structures most amplified by this process \citep{butler_threedimensional_1992}. Investigation of transient growth in Couette-Poiseuille flow has shown that adding even a small Couette component (introduced by a moving wall) to a Poiseuille flow (driven by a pressure gradient) significantly increases the nonmodal growth of the energy \citep{bergstrom_nonmodal_2005}. The flow thus becomes more sensitive to perturbations. As transient growth usually governs the dynamics of flow at early stages of transition to turbulence, one would expect Couette-Poiseuille flow to be less stable than pure Poiseuille flow.

Recall that linear transient growth cannot explain why turbulence does not decay for sufficiently high Reynolds number. A nonlinear cyclic process has been proposed that makes the turbulence self-sustained \citep{waleffe_self-sustaining_1997}. Instability of the streaks, manifested by their sinusoidal streamwise waviness \citep{duriez_self-sustaining_2009}, has been found to be necessary to maintain the turbulence. It has been shown quantitatively that the self-sustaining process is relevant to the evolution of turbulent spots in channel flow \citep{lemoult_turbulent_2014}.

We note that fully developed turbulence in Couette-Poiseuille flow has been studied numerically \citep{pirozzoli_large-scale_2011,bernardini_statistics_2011,gretler_calculation_1997,kuroda_direct_1995}, experimentally \citep{huey_plane_1974,stanislas_experimental_1992,telbany_velocity_1980,telbany_turbulence_1981,thurlow_experimental_2000,nakabayashi_similarity_2004} and theoretically \citep{lund_asymptotic_1980,wei_scaling_2007}. Specifically, the flow with zero net flux was investigated in Ref. \onlinecite{huey_plane_1974}. We complete our survey of Couette-Poiseuille flow by stating that a similar kind of flow profile appears in a long lid-driven cavity \citep{FLM1} or between two horizontal coaxial cylinders where the gap is partially filled with water (as in the case of Taylor-Dean instability in circular Couette-Poiseuille flow investigated in Ref. \onlinecite{mutabazi_oscillatory_1989,mutabazi_spatiotemporal_1990}.

Despite the studies cited above, plane Couette-Poiseuille flow has received relatively little attention until now, especially in the transitional regime. Having a relatively large test section and a base flow with zero mean advection velocity, we have been able to gain more insight into the dynamics of intermittent turbulent structures in shear flows. Until now, the only experimental attempts to generate stationary turbulent structures have been in plane Couette flow, which by definition has no streamwise pressure gradient. We present for the first time nearly stationary turbulent structures in a flow with non-zero streamwise pressure gradient, which represents a wide class of flows with practical relevance. Specifically, we report the first observations of turbulent spots localized in both the streamwise and spanwise directions. Another result of our research is the observation of the macro-organization of turbulence to form oblique turbulent bands in plane Couette-Poiseuille flow, as has been observed for Taylor-Couette \citep{coles_transition_1965}, Taylor-Dean \citep{mutabazi_spatiotemporal_1990}, plane Couette \citep{prigent_large-scale_2002,prigent_long-wavelength_2003,barkley_mean_2007,duguet_formation_2010,philip_temporal_2011} and plane Poiseuille flow \citep{tuckerman_turbulent-laminar_2014,tsukahara_experimental_2014}.

The article is organized as follows: in section \ref{sec:2} we describe our new experimental set-up. Next, in section \ref{sec:3}, we present a general characterization of our installation, including the natural transition to turbulence due to intrinsic noise of the facility. In section \ref{sec:4} we characterize the forced transition to turbulence which we triggered by applying a steady, continuous disturbance into the test section. Finally, in section \ref{sec:5} we discuss our results.

\section{Description of the experimental set-up}
\label{sec:2}

\begin{figure}
\begin{center}
\begin{minipage}[t]{0.495\linewidth}%[]{.5\linewidth}
\includegraphics[scale=1]{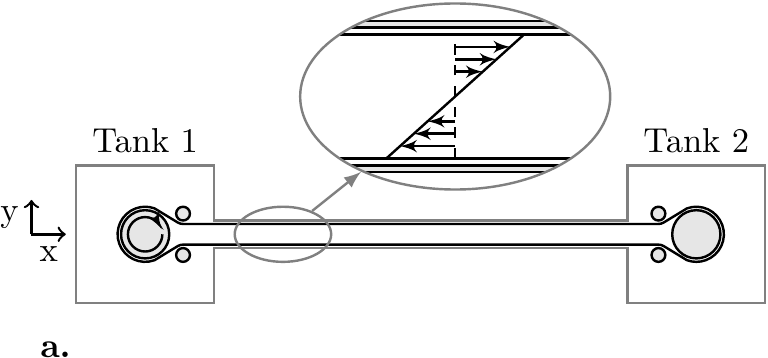}\\
\end{minipage}
\begin{minipage}[t]{0.495\linewidth}%[]{.5\linewidth}
\includegraphics[scale=1]{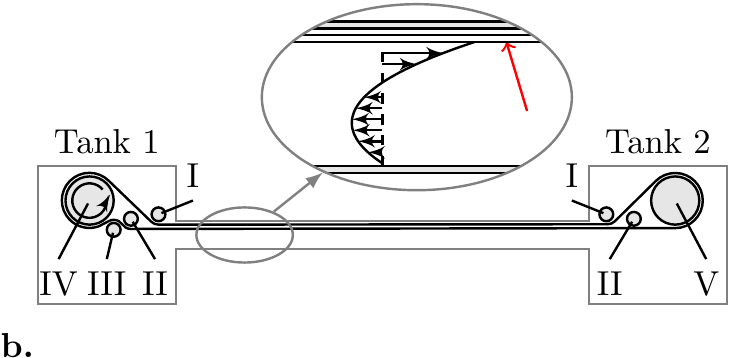}\\
\end{minipage}
\caption{Sketch of: a) the plane Couette experimental set-up \cite{tillmark_experiments_1992, daviaud_subcritical_1992}; b) our new facility to investigate plane Couette-Poiseuille flow. The upper moving wall induces the Couette flow to the right, which in turns generates the streamwise pressure gradient inducing Poiseuille flow to the left (compare with Fig.~\ref{fig:Scheme1}). The Roman numerals correspond to those of Fig.~\ref{fig:ExSetUp}. Red arrow in the inset marks the lower layer of the plastic belt.} 
\label{fig:1}      
\end{center}
\end{figure}

\begin{figure}[ht]
\includegraphics[scale=1]{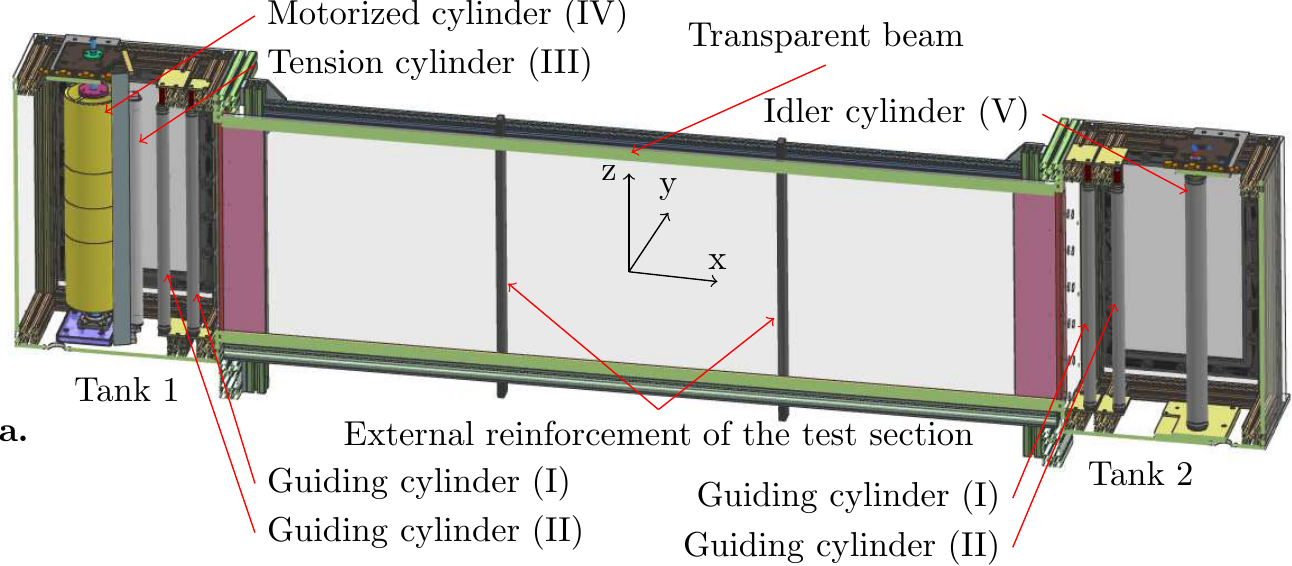}\\
\caption{Perspective view of the new experimental set-up with cross-section at the $y_*=0$ plane.}
\label{fig:ExSetUp}       
\end{figure}

\begin{figure}[ht]
\begin{minipage}[t]{0.48\linewidth}
\includegraphics[scale=1]{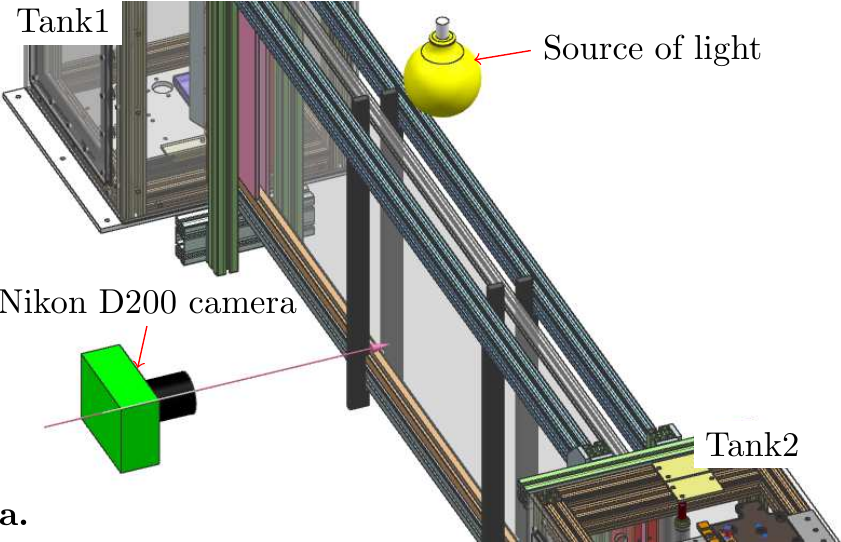} 
\end{minipage}
\begin{minipage}[t]{0.48\linewidth}
\includegraphics[scale=1]{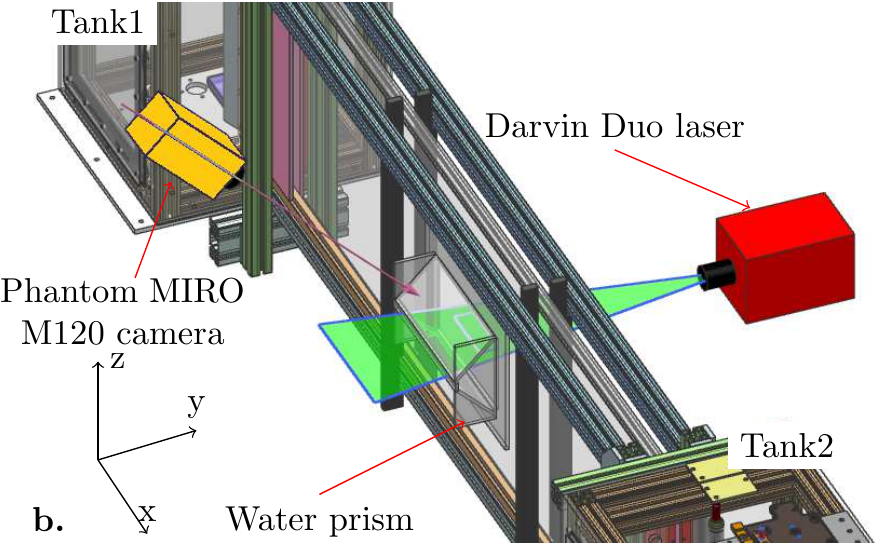}
\end{minipage}
\caption{Configuration we use to: a) perform flow visualizations. The source of conventional, incoherent light is placed at the top and the camera at the side of the test section; b) perform 2D PIV measurements.}
\label{fig:5}
\end{figure}

\begin{figure}[ht]
\includegraphics[scale=0.4]{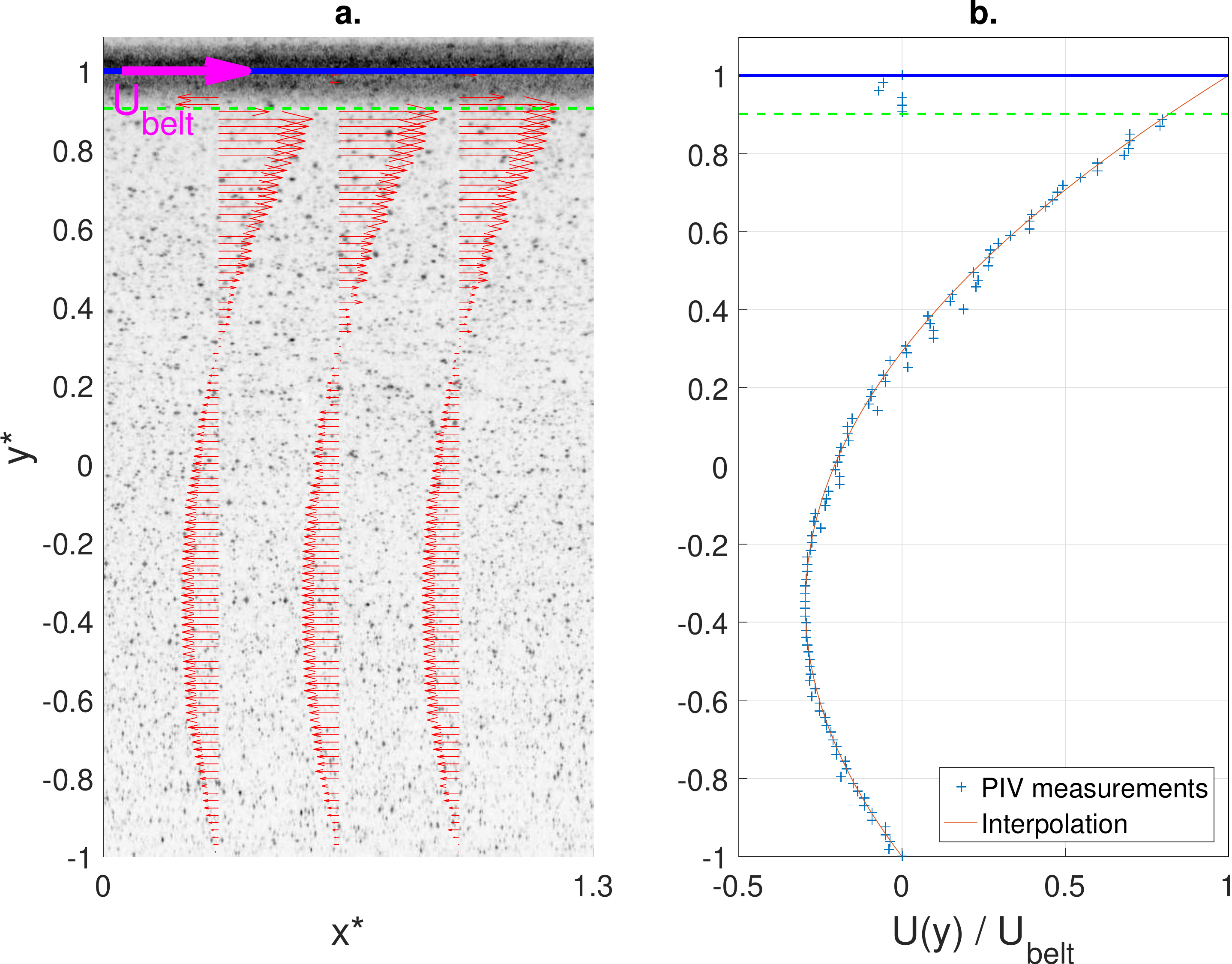}\\

\caption{a) Instantaneous snapshot of laminar flow with parabolic streamwise velocity profiles superposed on it. The velocity vectors are measured with the PIV technique by cross-correlating the particles seen in the photo. The stationary wall and moving belt are located at the bottom ($y_*=-1$) and top ($y_*=1$) respectively. Near the moving belt, we are not able to measure the velocity with PIV; b) representation of streamwise velocity profile $U_*(y_*)$ (blue crosses) as a function of wall-normal direction normalized with the belt speed. This example corresponds to the central velocity profile in a). We also show a quadratic interpolation of the measured velocity profile (solid red line), which can be observed to fit the data. The blue solid line at the top here and in a) represents the instantaneous $y_*$ position of the moving belt; the green dashed line marks the last point which can be measured with PIV. The interpolating function representing the velocity profile and the blue line cross very close to the $(U_*=U/U_{\rm belt}=1,y_*=1)$ point, which is precisely the position of the moving belt obtained from image processing.}
\label{fig:ExPiv}       
\end{figure}

\begin{figure}[ht]
  \includegraphics[scale=0.4]{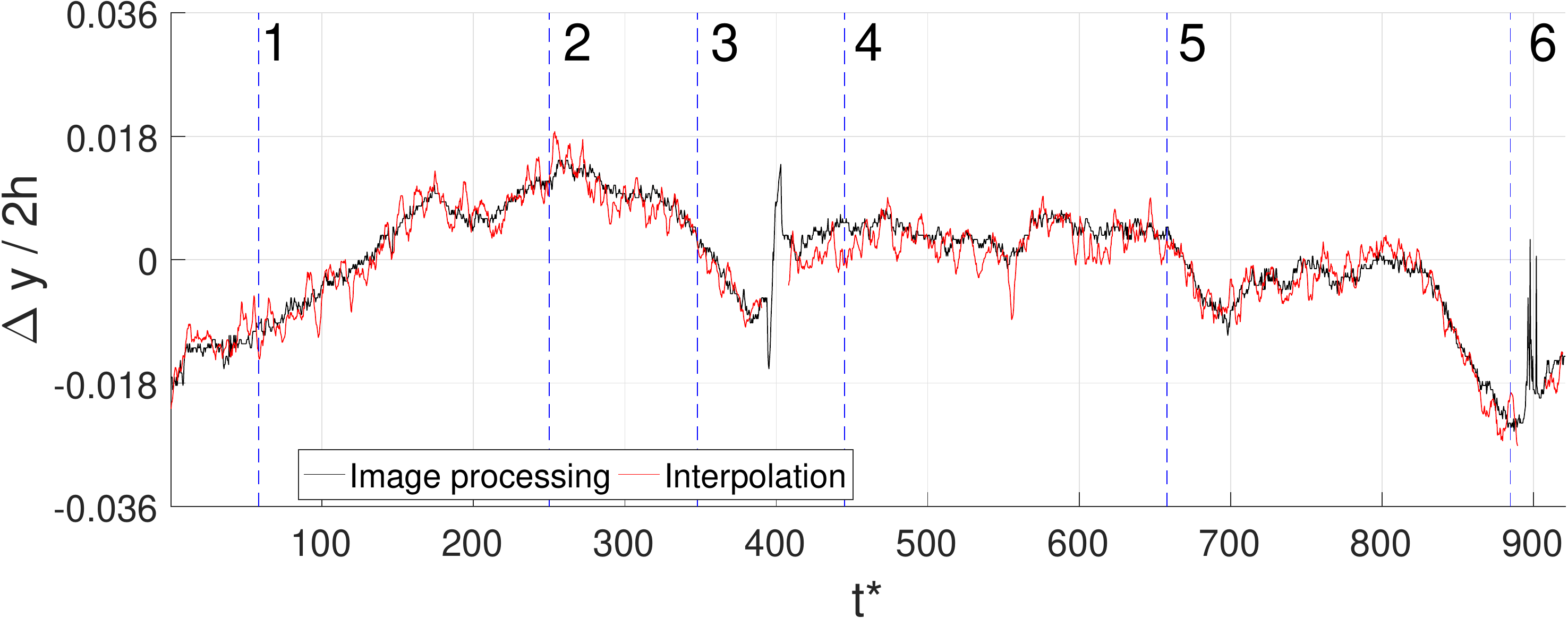}\\
\caption{Time series of deviations of the position of the moving belt. The ordinate represents the deviation from its time-averaged position ($y_*=1$). Black and red lines represent the belt position obtained with image treatment ($y_{\rm belt,img}$) and with interpolation ($y_{\rm belt,int}$) respectively. Two sharp peaks on the black lines at $t_*=400$ and $t_*=900$ are artifacts corresponding to the passage of the adhesive tape joining the two ends of the belt and do not represent the real film displacement. The vertical lines with numbers represent the instants at which instantaneous velocity profiles will be shown (see Fig.~\ref{fig:InterpolationProfiles}).}
\label{fig:TimeSeries}       
\end{figure}

\begin{figure}[ht]
  \includegraphics[scale=0.4]{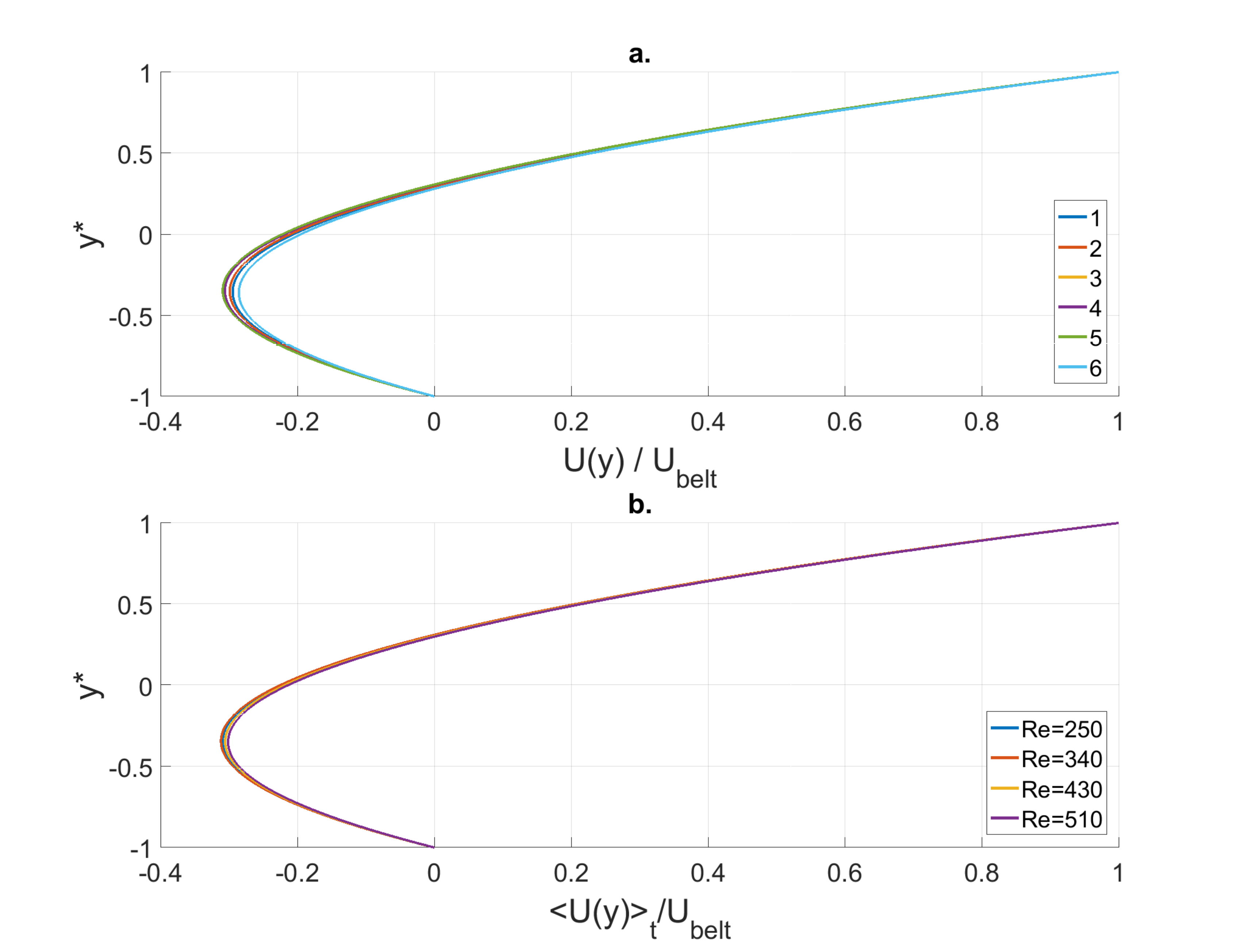}\\
\caption{a) Instantaneous streamwise velocity profiles obtained from interpolation of the measured streamwise velocity. The numbers corresponds to the instants marked by vertical lines in Fig.~\ref{fig:TimeSeries}; b) time-averaged velocity profiles $U_*(y_*)$ over one full period of the belt motion for different Reynolds numbers.}
\label{fig:InterpolationProfiles}       
\end{figure}

First, we denote by $x,y,z$ the streamwise, wall-normal and spanwise directions respectively. The center of the coordinate system is placed in the center of the test section. Our new installation is a generalization of the classical plane Couette experimental set-up (see Fig.~\ref{fig:1}a and Ref. \onlinecite{tillmark_experimental_1991,tillmark_experiments_1992,daviaud_subcritical_1992}). As shown in Fig.~\ref{fig:1}b we use looped plastic belt to impose the speed on one wall of the test section, while the other wall remains stationary. The moving wall drives the Couette flow toward the right side (red velocity profile on the left of Fig.~\ref{fig:Scheme1}), which in turn increases the pressure in tank 2. This positive streamwise pressure gradient induces the reverse Poiseuille flow (blue velocity profile in the center of Fig.~\ref{fig:Scheme1}). The resulting plane Couette-Poiseuille flow (black velocity profile on the right of Fig.~\ref{fig:Scheme1} and inset of Fig.~\ref{fig:1}b) is a superposition of these two contributions. It has zero mean advection velocity $\int_{-1}^{1} U(y)dy = 0$. We will also sometimes refer hereafter to the high-shear/low-shear regions close to the moving/stationary wall as the Couette/Poiseuille regions respectively.

In Fig.~\ref{fig:ExSetUp} we present a perspective view with a cross-section at the midgap plane to show in detail the side of the test section where the moving belt is placed. There are two lines of guiding cylinders (marked as I and II, see also Fig.~\ref{fig:1}b) which guide both layers of the plastic belt into the test section. We can regulate the $y$ position of both ends of these cylinders. There is one additional cylinder (III in Fig.~\ref{fig:1}, \ref{fig:ExSetUp}) to keep the plastic belt tight.  Its position (just after the motorized cylinder) was chosen carefully to provide us the best stabilization of the $z$ position of the belt when it moves. All the cylinders we use are provided by Interoll\textregistered \, with the exception of the motorized cylinder, which requires a large diameter (to diminish the slip between the cylinder and the moving belt) and a slightly tapered shape at both ends (to better control the $z$ position of the belt when it moves). For these reasons we have manufactured it on a three dimensional printer. Four additional external steel beams reinforce the test section and diminish the deflection of the side walls of the test section due to the hydrostatic pressure of the water.

The experimental set-up is mounted on a heavy granite table which provides mechanical stability and thermal inertia. The motorized cylinder is driven by a servo-motor produced by Yaskawa Electrics\textregistered \, ($100$ W) with a gear reduction of 1:26. We use glass plates (of $8$ mm thickness) as side walls, a Plexiglas beam as the upper wall and a transparent plastic belt made of Mylar\textregistered \,(of $175$ $\mu$m thickness), granting optical access to the test section. The gap between the glass walls of the test section is $2h_1 = 14$ mm. However, one can observe in schematic Fig.~\ref{fig:1}b that there are two layers of the plastic belt in the vicinity of the upper wall, which bounds the effective gap of the test section. We measure it with an optical method as $2h = 11.5$ mm. Hereafter we use the phrase moving belt to refer only to the lower (inner) layer of plastic film (indicated by the red arrow in the inset of Fig.~\ref{fig:1}b). The streamwise and spanwise dimensions of the test section are $2000$ mm and $540$ mm respectively. However, the width of the plastic belt is $520$ mm, which is our effective spanwise dimension. The belt is positioned slightly asymmetrically in the spanwise direction. Taking all of this into consideration, the aspect ratios of the test section in the streamwise/spanwise directions are $L_x/h = 347.8$ and $L_z/h = 90.4$ respectively. 

Spatial coordinates and time are nondimensionalized with the effective half gap $h$ and $h/U_{\rm belt}$ respectively and are marked by $*$ subscript. The velocity profile normalized by the belt speed is $$U_*=U/U_{\rm belt}=\frac{3}{4} ({y_*}^2-1)+\frac{1}{2}(y_* +1), ~~{\rm where}~~~ y_* = y/h \in (-1,1).$$ The Reynolds number is based on effective half gap $h$ (following the convention from plane Couette and plane Poiseuille flows) and the speed of the moving wall, $U_{\rm belt}$, namely $\Rey=U_{\rm belt} h/\nu$. We explore the range between $\Rey=160$ and $\Rey=780$.

We photograph the flow visualizations with a Nikon D200\textregistered \,camera (3800$\times$2800 pixels matrix) and Nikkor\textregistered \,$f=35$ mm lens. Its optical axis is collinear with the $y$ axis and the source of the white light is at the top of the test section (Fig.~\ref{fig:5}a). In addition, we also acquire supplementary video of flow visualization with video camera Canon\textregistered \, 3CCD XM2 Pal (720$\times$520 pixels), which will be described in details at the end of the section \ref{sec:4}. To perform the PIV measurements we use a Phantom MIRO M120\textregistered \,camera (1920x1600 pixels) with Nikkor\textregistered \,$f=85$ mm lens and a Darvin Duo\textregistered\, laser (double-headed, maximum output $80$ W, wavelength $527$ nm) in the configuration presented in Fig.~\ref{fig:5}b. We acquire a sequence of either single or double frame snapshots, which are then post-processed with Dantec Dynamic Studio\textregistered \,4.2 software.

We use a high concentration of small seeding particles made of Polyamid with a diameter of $5\mu$m. We use a rectangular 256 $\times$ 16 pixel interrogation window in the $x$ and $y$ directions with a 50\% overlap. This unconventional choice is justified by the dominant streamwise velocity component, which implies that the streamwise pixel displacement is an order of magnitude larger than that in the wall-normal direction. The rectangular window enables us to increase the signal-to-noise ratio, keeping a high spatial resolution in the $y$ wall-normal direction. With this procedure we measure instantaneously three velocity profiles with $0.3h$ spacing in $x$ and with 100 points across the gap.

When the camera is placed on top of the test section (with its optical axis aligned along the $z$ axis), we can measure the streamwise velocity with PIV only in the vicinity of the upper wall of the test section due to the high concentration of seeding particles. In order to measure the streamwise velocity profiles at different spanwise locations, we put the camera on the side of the test section (with its optical axis inclined at $45^{\circ}$ with respect to the $z$ axis, see Fig.~\ref{fig:5}b). We also use a Scheimpflug mount to record a well-focused image despite the inclination of the camera, as well as a water prism to reduce the optical distortions due to the difference in refractive indices of water and air.

In order to study the laminar base flow, we measure the instantaneous streamwise velocity for different Reynolds numbers ($\Rey \in (250,340,430,510)$) in the central part of the test section. We acquire a single image sequence and correlate two consecutive images. We set a high enough frequency (from $19$ Hz to $42$ Hz, depending on Reynolds number) to retain the time correlation between two snapshots. We need to record about 2800 images on the 3 GB internal memory of the Phantom\textregistered \, camera to cover one period of the belt motion. For this reason we use part of the camera matrix (512 pixels in $x$ and 896 pixels in $y$). This procedure provides us with the best possible temporal resolution for a given spatial resolution (directly related to the size of the camera matrix) and for a given measurement time. In Fig.~\ref{fig:ExPiv}a we present one example of an instantaneous PIV vector field acquired for $\Rey=510$, which shows three similar velocity profiles within the measurement area. We plot the central profile in $x$ in Fig.~\ref{fig:ExPiv}b.
 
Fig.~\ref{fig:ExPiv}a shows that the width of the optical image of the belt is about 1 mm, which is more than its actual thickness. This is a consequence of the inclination of the optical axis of the camera with respect to the laser sheet of finite width. Indeed, the light coming from the laser sheet of finite width is reflected by the belt and then registered on the camera matrix as a thick line. There are additional contributions from the defocusing and scattering of the laser light in the vicinity of the moving belt. In the region above the dashed green line we are not able to measure the velocity with our PIV technique, as it produces many spurious vectors. We define the center of this thick line as the instantaneous position of the moving belt ($y_{* \rm belt,img}$) and we determine it for each image using edge detection techniques. In Fig.~\ref{fig:ExPiv} we mark this instantaneous belt position by a solid blue line and we assign the value $y_*=1$ to this location. In Fig.~\ref{fig:TimeSeries} we present as a solid black line a time series of the deviations of the moving belt position from its time-averaged location. The actual belt position changes smoothly in time.

As in the measured data in Fig.~\ref{fig:ExPiv}b there is a maximum and due to the fact that we expect the laminar Couette-Poiseuille flow to be a quadratic function of $y_*$, we interpolate the measured velocity points using a quadratic polynomial of the form $U_*(y_*)=\sigma_1(y_*^2-1) + \sigma_2(y_*+1)$ (red line in Fig.~\ref{fig:ExPiv}b). The interpolation fits the data very well. Then we estimate the $y_{*\rm belt,int}$ position at which the interpolation function reaches the known value of belt speed ($U(y_{*\rm belt,int})=U_{\rm belt}$). In Fig.~\ref{fig:ExPiv}b the interpolated streamwise velocity profile (red line) and the measured belt position (thick blue line) cross very close to the point $(1,1)$ as expected, which confirms the validity of our interpolation curve. We also compare the time evolution of the two wall positions obtained by these two methods ($y_{*\rm belt,int}$ as the red and $y_{*\rm belt,img}$ as the black line). The belt position predicted by interpolation matches very well the real position of the moving belt, with a deviation of less than 0.1 mm. We note that this is the first time that a detailed study of the gap variation is performed for this type of experiment with moving walls (including plane Couette facilities).

In order to check whether the base flow is affected by temporal fluctuations of the moving belt position we plot in Fig.~\ref{fig:InterpolationProfiles}a instantaneous interpolations of the streamwise velocity profiles for six different instants, which are marked in Fig.~\ref{fig:TimeSeries} by dashed vertical lines and numbers from 1 to 6. These velocity profiles are virtually the same, which proves that the base flow does not depend in a significant way on the phase of the belt motion. We also calculate the time averaged velocity profiles $<U(y_*)>_t$ for different Reynolds numbers (Fig.~\ref{fig:InterpolationProfiles}b). They collapse onto a single curve after being normalized with the belt speed. Finally, we calculate the mean advection velocity of the time averaged velocity profile ($U_{\rm avg}=\frac{1}{2h}\int_{-1}^1 <U(y_*)>_tdy$), which does not exceed $0.03 U_{\rm belt}$ in the central part of the test section.

Having determined the fluctuations of the belt position (Fig.~\ref{fig:TimeSeries}), as well as the variation of the velocity profiles in time (Fig.~\ref{fig:InterpolationProfiles}a) and for a given Reynolds number (Fig.~\ref{fig:InterpolationProfiles}b), we can estimate the total error of the local Reynolds number (for a given $z$ position) as lower than 5\%. The variation of the effective gap for different $z_*$ locations is lower than $\pm 0.5$ mm. The spatial variation of the temperature in the test section does not exceed $0.2^{\circ}$C. The resulting error related to the fluid viscosity is less than $0.6\%$. We estimate the global variation of the Reynolds number for different $z_*$ locations as lower than $7.6\%$. The cross-flow component of the base flow (in the spanwise direction) is lower than $2.0\%$.

In all subsequent figures the direction of motion of the plastic belt is toward the right.

\section{Characterization of the natural transition to turbulence triggered by the intrinsic noise of the installation}
\label{sec:3}

\begin{figure}[ht]
\includegraphics[width=\linewidth]{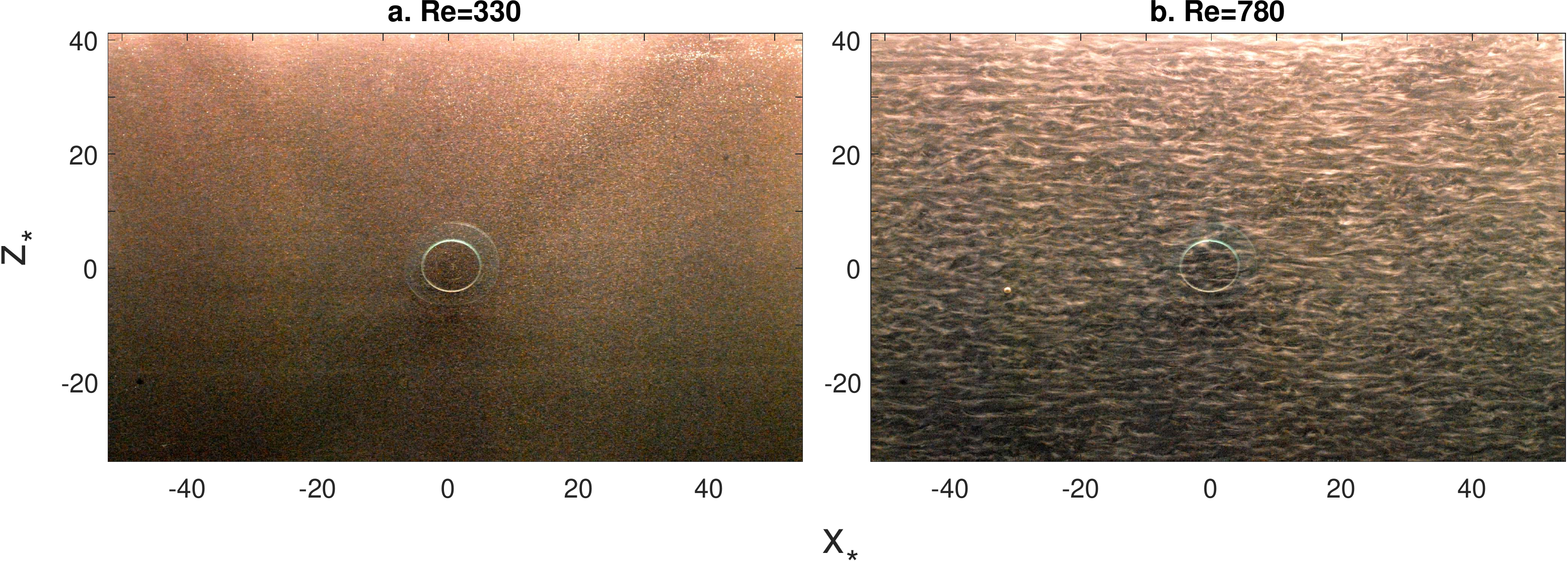}
%\begin{tikzpicture}
%
%	\node[anchor=south west,inner sep=0] (image) at (0,0) {\includegraphics[width=\linewidth]{Fig8.eps}};%
%	\draw[->,red] (1.5,1.25) -- node[anchor=south]{$U_{belt}$} (2.5,1.25)  ;
%\end{tikzpicture}
\caption{Flow visualizations for different Reynolds numbers: a) uniform laminar flow ($\Rey=330$); b) featureless turbulent region in the entire test section ($\Rey=780$).}
\label{fig:2}       
\end{figure}

\begin{figure}[ht]
  \includegraphics[width=\linewidth]{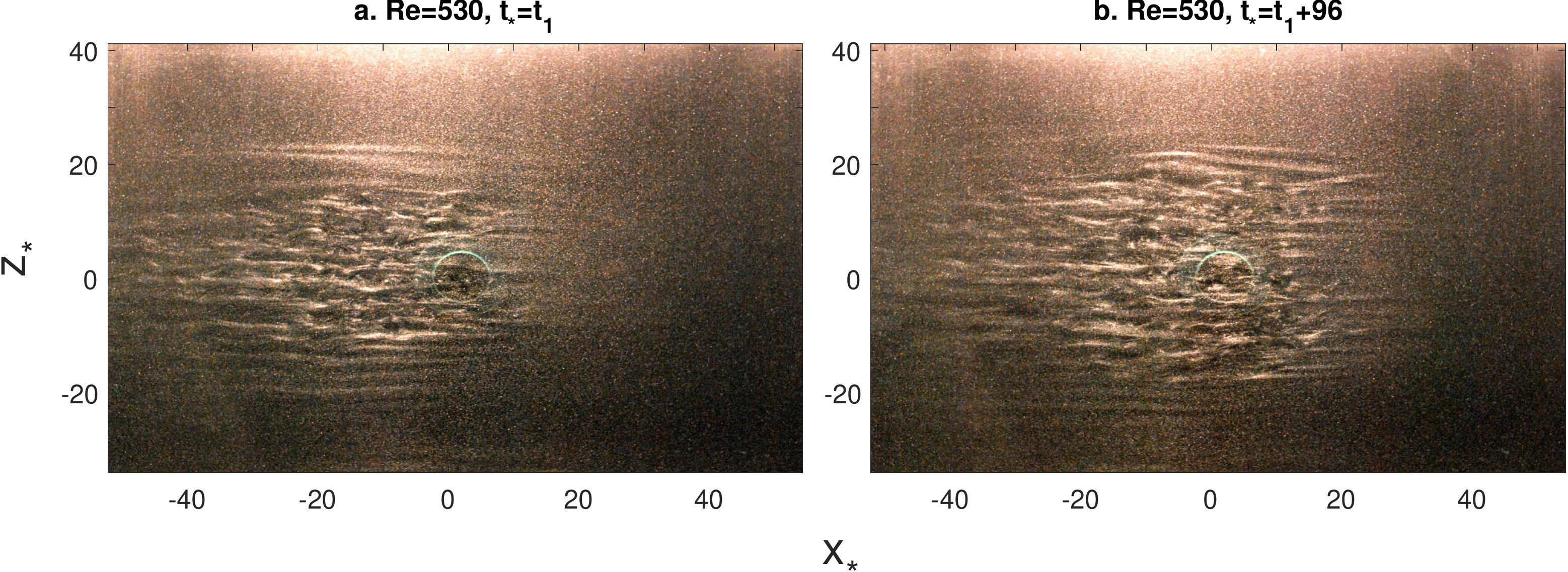}\\%[width=8cm]
\caption{Flow visualizations of a localized turbulent spot surrounded by laminar flow in plane Couette-Poiseuille flow ($\Rey=530$). Sequence of pictures shows the slow advection of the turbulent structure toward the right (with time interval of $96$ advection units).} 
\label{fig:4}      
\end{figure}

\begin{figure}[ht]
  \includegraphics[width=\linewidth]{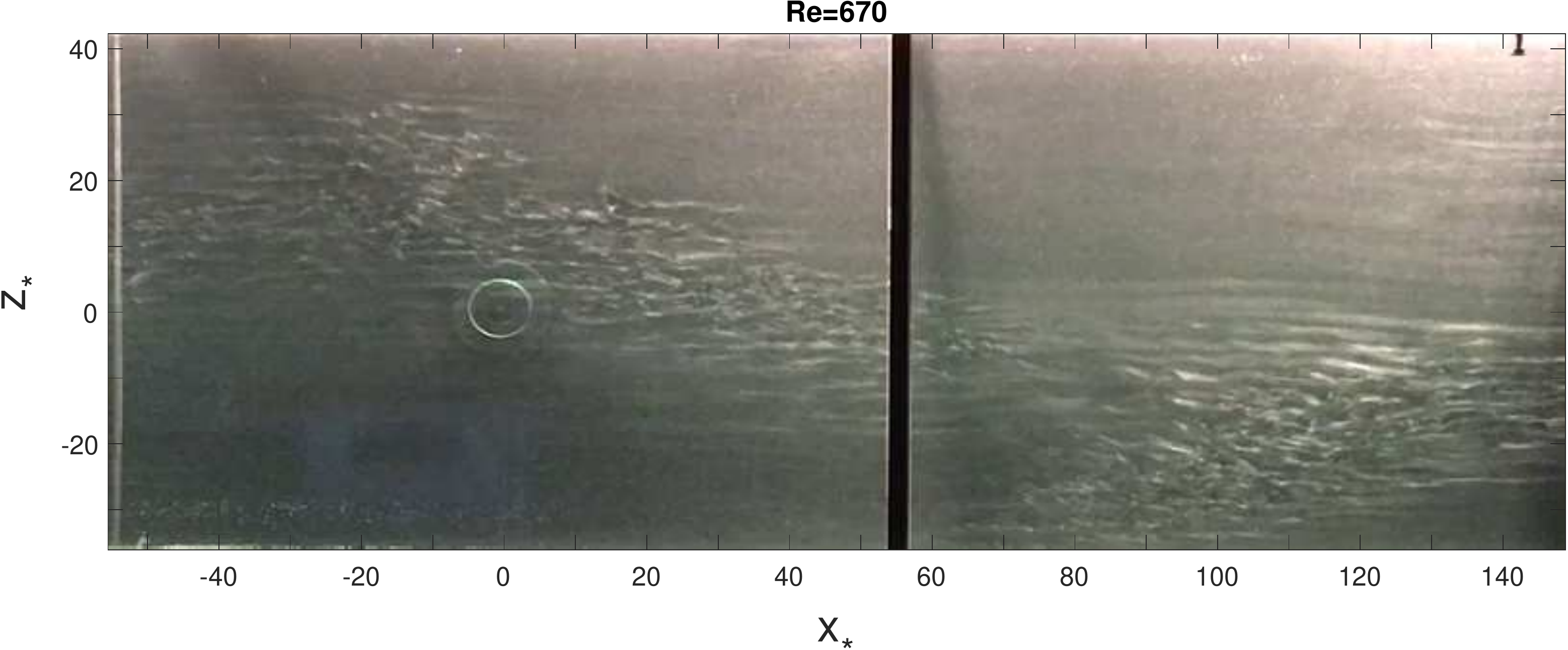}\\
\caption{Example of macro-organization of turbulent spots in the form of oblique turbulent bands ($\Rey=670$).}
\label{fig:3}      
\end{figure}
 
\begin{figure}[ht]
  \includegraphics[scale=0.28]{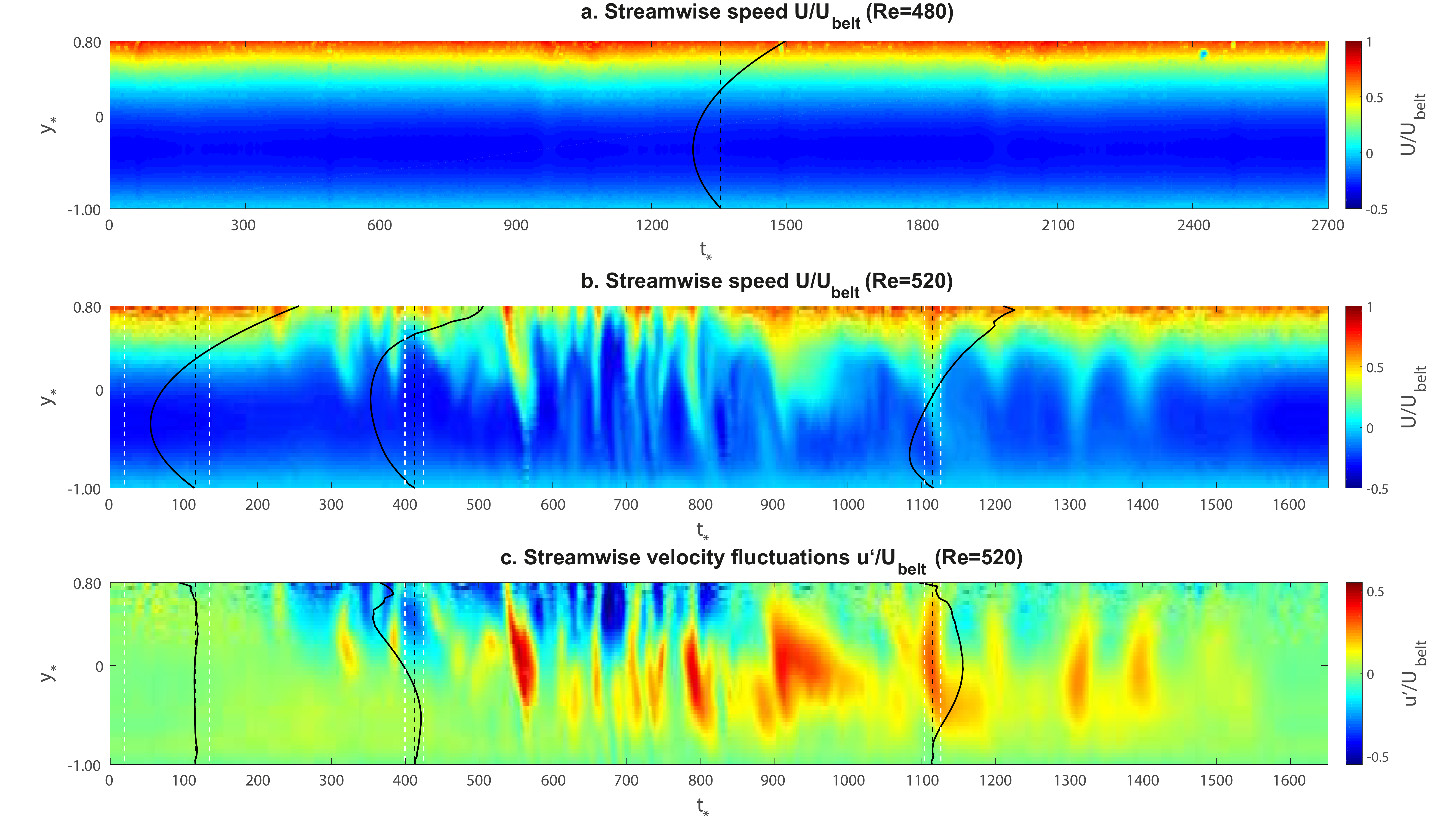}\\
\caption{Spatio-temporal diagram of: a) measured streamwise velocity, $U_*(t,y_*)$, for the laminar flow ($\Rey=480$). The black solid line represents the time-averaged profile; b) measured streamwise velocity, $U(t,y_*)$, for the intermittent flow ($\Rey=520$); c) streamwise velocity fluctuations $u_*'(t,y_*)=(U(t,y_*)-U_{\rm base}(y_*))/U_{\rm belt}$. We consider the time-averaged streamwise velocity profile in the range $t_* \in (20-120)$ as the base flow without perturbations. The pattern at $t_* \in (800, 1400)$ is a signature of the unsteady, wavy structure of the turbulent spot. The black profiles have been calculated by time-averaging the instantaneous velocity profiles within the ranges marked by white dashed lines.}
\label{fig:STturb}       
\end{figure}

We perform flow visualizations to characterize qualitatively the flow in the test section. For this purpose we use reflective aluminium flakes (STAPA IL HYDROLAN 2154 55900/G produced by ECKART) of typical diameter $d_{al}\in(30$ $\mu$m$,80$ $\mu$m$)$, which are dispersed in water. These tracers enable the detection of three dimensional vortical structures in the flow, through spatial fluctuations of reflected light intensity. In contrast, the light distribution in laminar regions is nearly uniform and featureless. In this way, turbulent regions can be distinguished. The pictures presented here are taken with a Nikon camera with a 3800$\times$ 2800 pixel matrix, with a pixel pitch equal to $0.167$ mm.

Note that the fluid in the main tanks is continually disturbed by the rotating cylinders, which makes it always turbulent for the range of Reynolds numbers considered here. For this reason the inlets of the test section are the sources of the natural perturbations. Even though these operating conditions are similar to those presented in Ref. \onlinecite{tsukahara_experimental_2014} for Poiseuille flow, we recall here that the mean advection velocity is nearly zero, so the turbulent flow is not advected from the inlets to the test section.

The flow is laminar in the entire test section up to $\Rey \simeq 420$ (as in Fig.~\ref{fig:2}a). For higher Reynolds numbers some turbulent structures appear at the left entry ($x_*<0$) generated by the turbulence in the main tank. Up to $\Rey \simeq 480$, the amplitude of these perturbations is not strong enough to trigger the transition and the turbulent structures present in the main tank do not propagate further into the test section.

For $\Rey \gtrsim 480$ the Couette-Poiseuille flow is no longer stable and the test section is occasionally invaded by transient patches of turbulence. However, up to $\Rey \simeq 510$ these events rarely occur and the undisturbed laminar base flow can persist for most of the time. In figure \ref{fig:4}a,b we present a sequence of images illustrating such a localized turbulent spot surrounded by laminar flow, which is slowly advected to the right with a very small advection speed of $U_{\rm advection}\simeq 0.095 U_{\rm belt}$.

As we increase the Reynolds number even further (to $\Rey \simeq 670$), the spots expand obliquely to form a turbulent structure reminiscent of laminar-turbulent bands, one example of which is presented in Fig.~\ref{fig:3}. Finally, at high enough Reynolds number ($\Rey \simeq 780$), the flow is uniformly turbulent (Fig.~\ref{fig:2}b).

In order to demonstrate the transition to turbulence in more detail, we measure the instantaneous streamwise velocity as a function of $y_*$ with the PIV configuration shown in Fig.~\ref{fig:5}. We acquire double-frame images with the sampling frequency of 2 Hz, which are cross-correlated to determine the instantaneous velocity fields. In Fig.~\ref{fig:STturb}a we present the spatio-temporal diagram of a single streamwise velocity profile $U_*(t_*,y_*)$ for low Reynolds number ($\Rey=480$) measured at $(x_*=0,z_*=0)$. The isocontours on this diagram are nearly horizontal, which shows that the flow is laminar and does not depend on time. This corresponds to the visualization of a laminar flow shown in Fig.~\ref{fig:2}a). We determine the base flow profile $U_{\rm base}(y_*)$ by time-averaging the results within the entire sequence of measurements. The resulting profile (black solid line in the center of Fig.~\ref{fig:STturb}a, see also Fig.~\ref{fig:InterpolationProfiles}b) is a quadratic polynomial.

As we increase the Reynolds number to $\Rey = 520$, we observe a transition to turbulence triggered by intrinsic noise of the installation. In Fig.~\ref{fig:STturb}b the flow becomes locally time-dependent/intermittent for $t_* \in (170, 1550)$, due to the passage of the localized turbulent spot through the PIV measurement section (compare also with the visualizations of the turbulent spot in Fig.~\ref{fig:4}). However, for $t_* \in (0, 170)$ the flow is stationary and laminar. We calculate the time-averaged profile in this range (parabolic profile on left side of Fig.~\ref{fig:STturb}b), which we consider as the base flow without perturbation $U_{\rm base}(y_*)$. Then we subtract it from the measured instantaneous streamwise velocity component $U(t,y_*)$ to calculate the streamwise velocity fluctuations $u'(t,y_*)=U(t,y_*)-U_{\rm base}(y_*)$ (Fig.~\ref{fig:STturb}c).

We can clearly observe the unsteady structure of these fluctuations, a signature of a turbulent spot in plane Couette-Poiseuille flow. The transition starts in the vicinity of the moving wall, in the high shear region ($t_* \in (250, 500)$), and then, as the turbulent structure grows, it gradually spreads across the whole gap. Finally the flow relaxes back to the laminar state ($t_*>1550$).

We also show two examples of turbulent streamwise velocity and fluctuation profiles ($t_*=415$ and $t_*=1100$ in Fig.~\ref{fig:STturb}b,c), which are calculated by time averaging the data within the range delimited by white dashed lines. The profile at $t_*=415$ shows the wall-normal transfer of fluid with negative velocity from the low-shear region toward the high-shear region, whereas the profile at $t_*=1100$ shows the opposite. Note in Fig.~\ref{fig:STturb}c that near the moving wall, the streamwise velocity fluctuations are negative, whereas away from the wall (low shear region) the streamwise velocity fluctuations are positive.

These PIV measurements demonstrate that plane Couette-Poiseuille flow can be also regarded as asymmetric Poiseuille flow with one active, high shear region near the moving belt. This is in contrast to the classical symmetric Poiseuille profile with two active regions, one next to each wall.

\section{Transition to turbulence triggered by a permanent perturbation}
\label{sec:4}

\begin{figure}[ht]
  \includegraphics[scale=0.68]{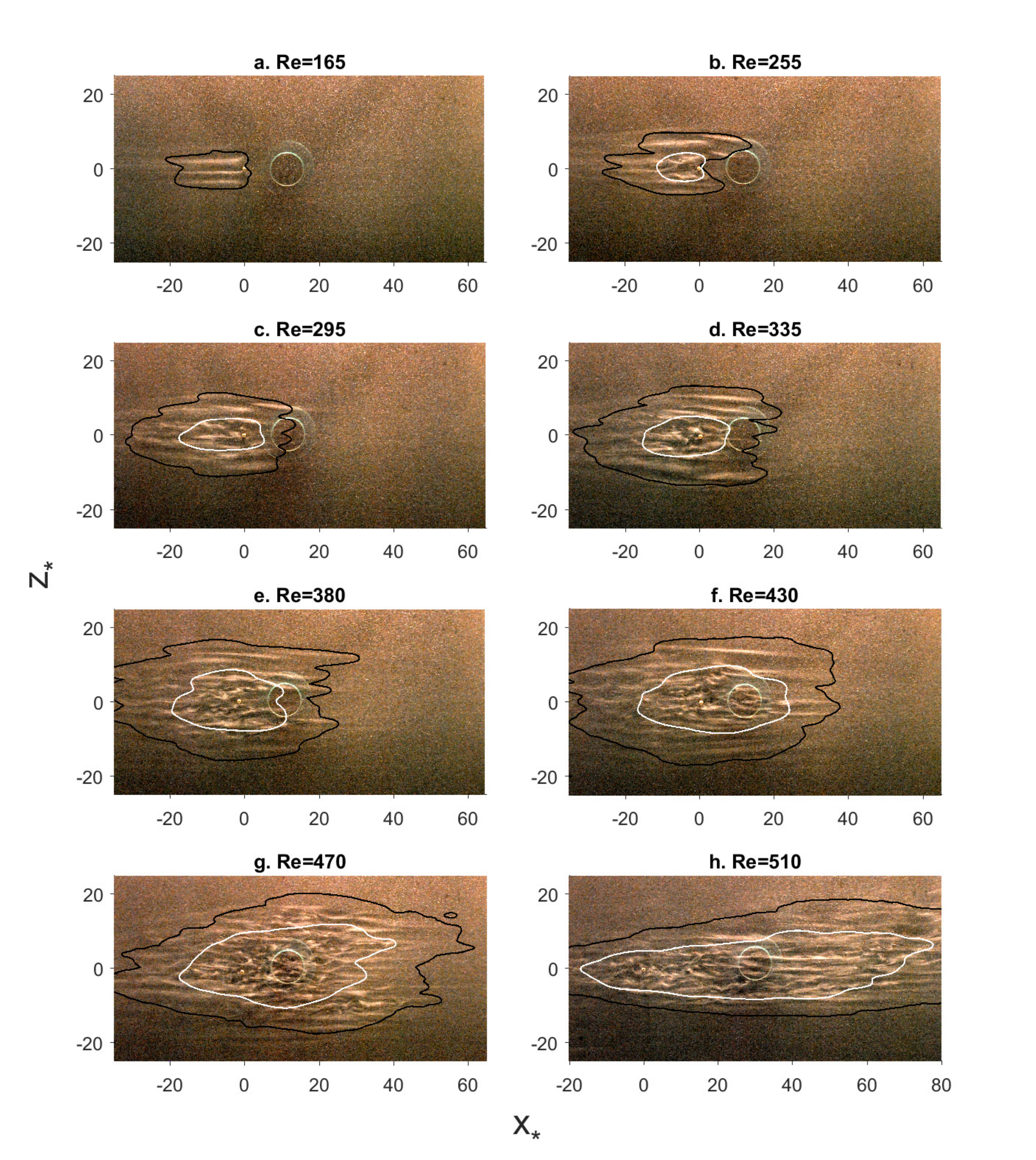}\\
\caption{Instantaneous flow visualizations for different Reynolds numbers. The transition to turbulence is triggered by a sphere placed in the test section near the moving wall. The $(x_*,z_*)$ origin is now located at the center of the sphere. The stationary/moving wall is closer to/further from the reader and the direction of the moving wall is toward the right. We also superpose on these images the time-averaged envelope contours that represent the total area of the spot (black contours) and its active turbulent core (white contours).}
\label{fig:FT_Vis8a}       
\end{figure}

\begin{figure}[ht]
  \includegraphics[scale=0.5]{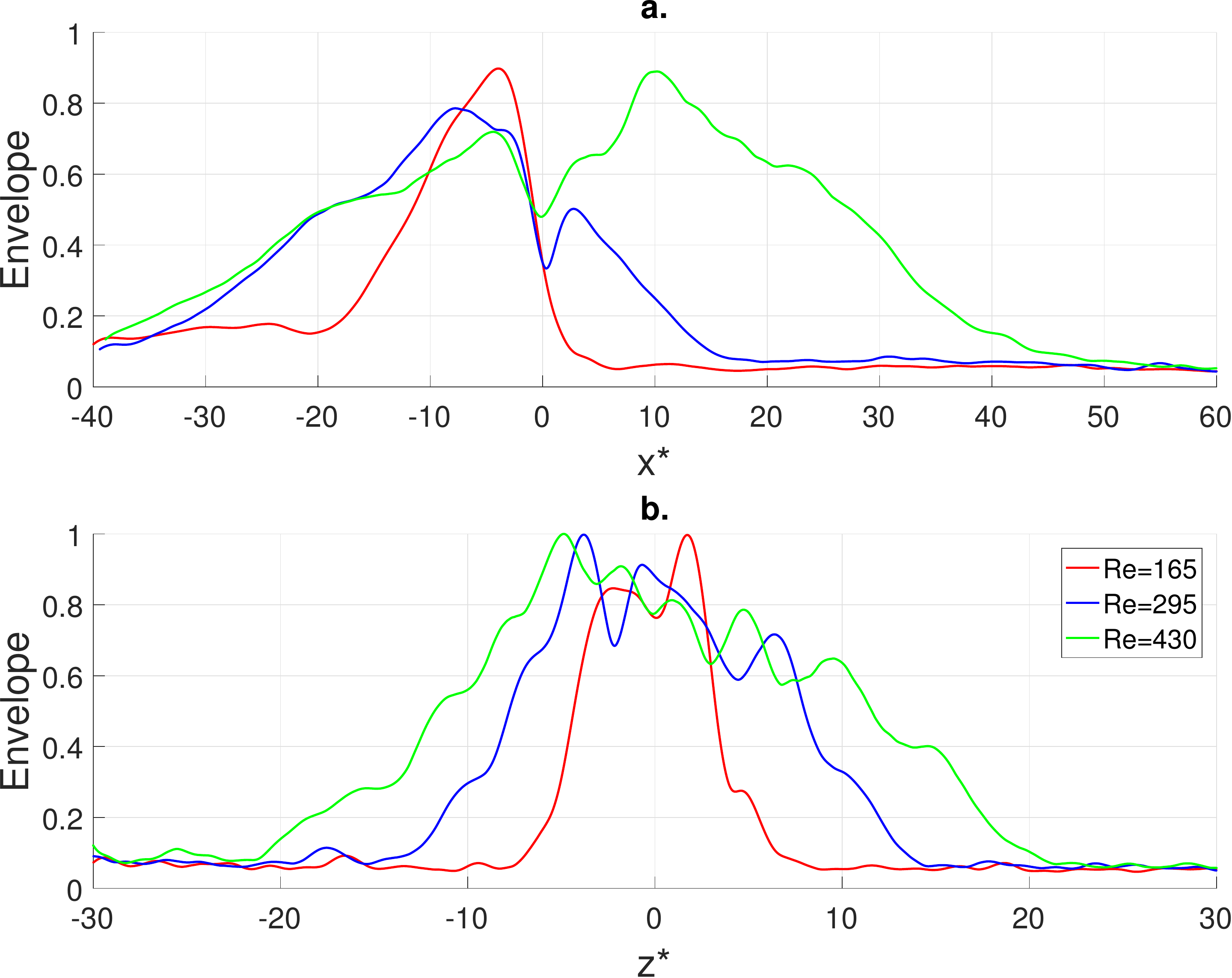}\\
\caption{Time-averaged envelope representing the total area of the turbulent spot: a) spatial extent of the envelope along the streamwise direction for $z=0$; b) spanwise extent of the envelope, where the $x$-coordinate has been selected for each Reynolds number to maximize the spanwise extent.}
\label{fig:ENV_RES}       
\end{figure}

\begin{figure}[ht]
  \includegraphics[scale=0.5]{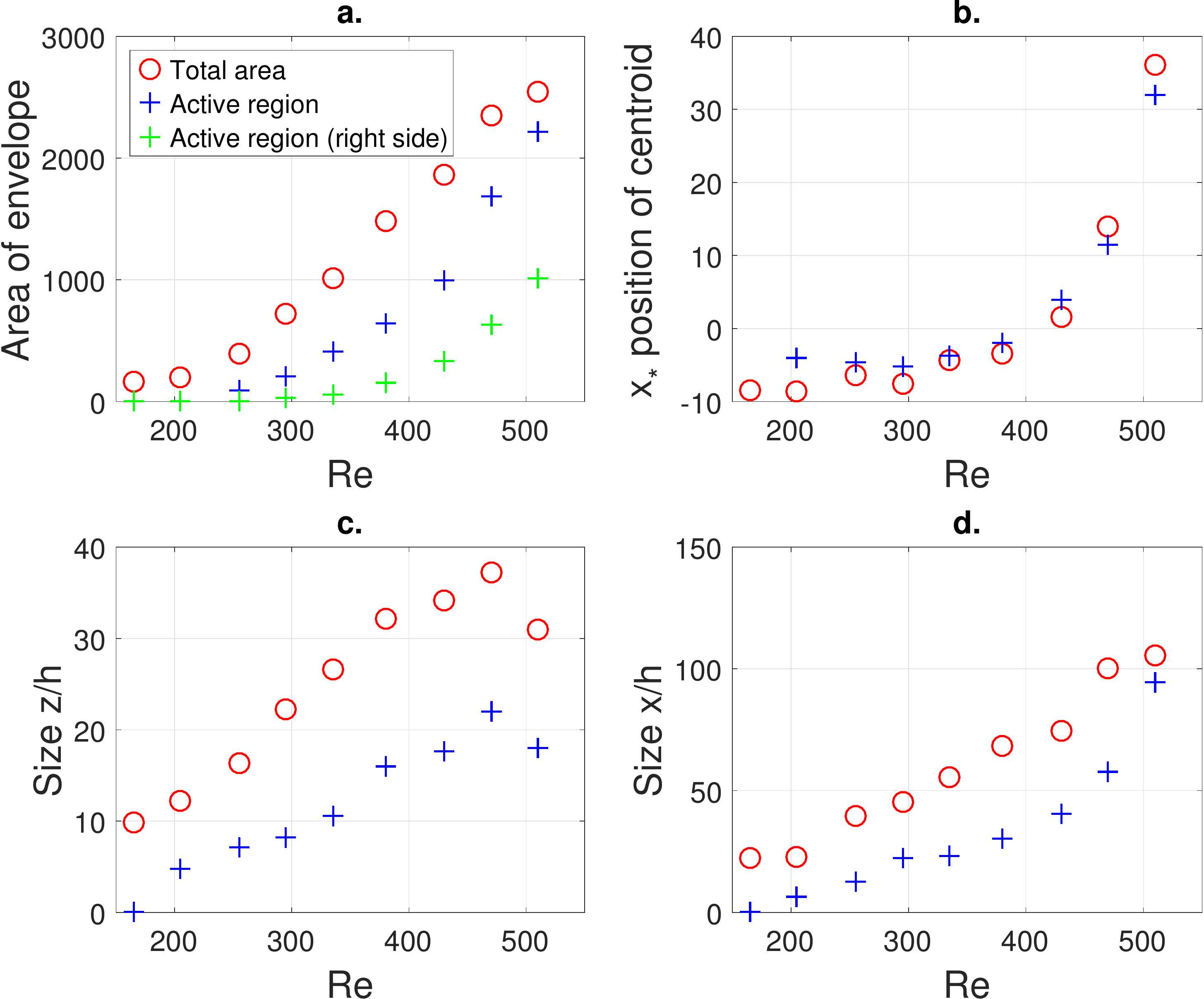}\\
\caption{Dependence on $\Rey$ of the envelopes of the total area and the active core of the turbulent spot: a) the area; b) the $x_*$ position of the centroids; c),d) the extent in $z$ and $x$ directions.  Crosses correspond to the active area (containing wavy streaks), while circles represent the region which includes in addition the weak undisturbed streaks and oblique waves at the laminar-turbulent interface.}
\label{fig:ENVplots}       
\end{figure}

\begin{figure}[ht]
  \includegraphics[scale=0.65]{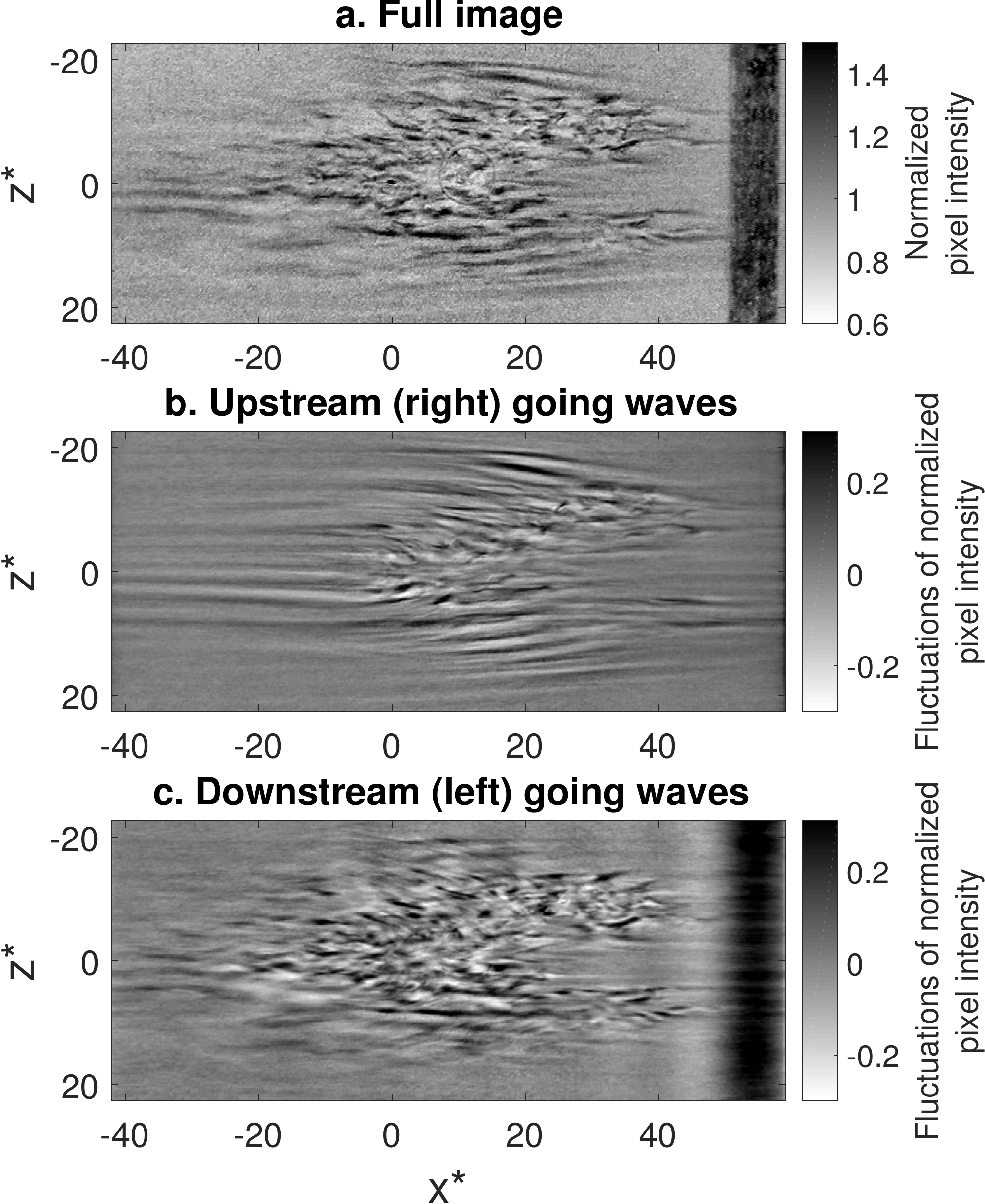}\\
\caption{Decomposition of the flow visualization of the turbulent spot
  into the structures that move upstream and downstream: a) full image
  of flow visualization; b) pattern which moves upstream (right); c) pattern which moves downstream (left). All three pictures correspond to the same instant of time.}
\label{fig:LeftRight}       
\end{figure}

In this section we present flow visualizations for the transition forced by an external and permanent perturbation. For this we insert a ferromagnetic sphere of diameter $6.2$ mm, which is held at a fixed position within the test section by a strong magnet. The sphere touches the moving wall and so is within the high shear region. In addition, the friction with the moving belt causes rotation of the obstacle. However, this imposed frequency is higher than the typical frequencies observed in the flow and thus can be neglected. This obstacle locally modifies the flow \citep{bottin_discontinuous_1998}, creating a steady, localized disturbance.

For each Reynolds number, we take a sequence of 90 images with sampling frequency $f=1$ Hz. We recall that we acquire the images with very high spatial resolution (3800 $\times$ 2800 pixel matrix). We shift the origin of the coordinate system with respect to Fig.~\ref{fig:2}-\ref{fig:4} by placing it at the center of the sphere. We call the left/right side of the sphere the downstream/upstream direction, taking the direction of the back flow (Poiseuille component) as the reference. In Fig.~\ref{fig:FT_Vis8a} we present flow visualizations representing the flow structure as the Reynolds number is increased. For $\Rey = 165$ we observe a few stationary streamwise vortices which expand towards the left (Fig.~\ref{fig:FT_Vis8a}a). The vortical structure observed on the left side of the sphere is probably due to a pair of streamwise counter-rotating vortices generated in the wake of the sphere, which, in uniform background flow, appears at $\Rey_{\rm sphere}\simeq 210$ \citep{johnson_flow_1999,gumowski_transition_2008}, where $\Rey_{\rm sphere}=(U_{\rm freestream}\,d_{\rm sphere})/\nu$. In our case $d_{\rm sphere}=6.2$ mm $\simeq h$, which implies that $\Rey_{\rm sphere} \simeq \Rey$. For $\Rey = 255$ the turbulence starts to invade the right side of the sphere (note the appearance of small vortices for $x_*>0$ in Fig.~\ref{fig:FT_Vis8a}c). As the Reynolds number is further increased, this streamwise extent to the right becomes increasingly important.

We have also observed that for $\Rey \lesssim 480$ the spot stays in a fixed location pinned to the sphere, but for higher Reynolds numbers the size of the spot fluctuates and it moves toward the right. This can be compared with the front speeds in pipes for puffs, where the upstream front (with respect to the direction of the Poiseuille component) travels more slowly downstream than the average velocity of the base flow \citep{barkley_rise_2015}. The analogue of this situation in our case is the motion to the right (upstream).

The spots have a preferred inner structure (a spanwise-periodic pattern of streamwise streaks) with wavelength $\lambda_z$ about $2.5h$ (representing the wave vector ($k_{x*}=0, k_{z*}=2.52$) in Fourier space). However the spot structure also includes oblique waves (i.e. straight streaks which are oriented slightly obliquely with respect to the streamwise direction) at the laminar-turbulent interface and undulated (or wavy) rolls in the center of the spot, which broaden the spatial Fourier spectrum.

In order to describe the dependence of the area of the turbulent spot on Reynolds number, we use the two-dimensional Hilbert transform to compute the envelope of the modulated function of gray levels representing the spot. First, we normalize the pixel intensity of each image by dividing it by the background reference corresponding to the laminar flow without the sphere. Then we compute its two dimensional FFT spectrum and we filter it, retaining the range $|k_{x*}| < 1.19$ and $k_{z*} \in (0.76, 4.28)$. Next, we use the two-dimensional inverse FFT transform to compute the filtered spot and we get its envelope/amplitude $\rho(x,z)$. Finally, we compute for each Reynolds number the time-averaged spatial envelope for all images in the sequence. This is justified because the global dynamics of the turbulent spot is nearly stationary, as the forcing is constant in time and the turbulent region is pinned to the sphere until $\Rey \simeq 480$.

We also estimate the size of the more dynamically active, turbulent region at the core of the turbulent spot. To do this we use the observation that this region is directly related to streamwise waviness of the streaks, which resembles travelling waves. The streamwise dependence of wavy streaks is manifested by the appearance of modes with $k_{x*} \neq 0$ in the spatial spectrum. However, such modes generate higher harmonics. This effect is further increased by the fact that we analyse the pixel intensity of flow visualizations, which adds spurious nonlinear content. As a result we are not able to identify a single mode which corresponds to the streak waviness. Instead, we consider the spectral range $|k_{x*}| \in (1.13,1.85)$ and $k_{z*} \in (1.60,3.44)$, which is related to the harmonics of this structure. In this way, we can insure that the envelope computed from this spectral region corresponds to the short wavelength streamwise undulation rather than to the long oblique straight waves. In order to describe a spatial distribution of both regions (namely the active core related to the waviness of the streaks and the total area, which in addition includes the surrounding region with oblique waves at the laminar-turbulent interface) for different Reynolds numbers, we superpose iso-contours of both envelopes on the flow visualization pictures (Fig.~\ref{fig:FT_Vis8a}). Note in Fig.~\ref{fig:FT_Vis8a}a that there is no active region for $\Rey = 165$, since the structures there are wake vortices generated by the sphere and unrelated to turbulence.

In order to better illustrate how the size of the turbulent spot changes with increasing Reynolds number, we show spatial profiles of the total spot envelope along the $x_*$ (Fig.~\ref{fig:ENV_RES}a) and $z_* $ (Fig.~\ref{fig:ENV_RES}b) directions. The former is plotted for $z_*=0$, and the latter for the value of $x_*$ which maximizes the size along the $z_*$ direction. For low Reynolds numbers ($\Rey=165$ in Fig.~\ref{fig:FT_Vis8a}a and \ref{fig:ENV_RES}a) all of the activity takes place on the left side of the sphere. The upstream front is steep, whereas in the downstream direction the envelope slowly decays to zero with a large tail extending toward $x_*<0$. As we increase the Reynolds number the turbulent region extends further and further upstream.

In Fig.~\ref{fig:ENVplots} we present several quantities to further characterize this dependence. Fig.~\ref{fig:ENVplots}a shows that the area of both the total and the active regions increase monotonically with Reynolds number. In Fig.~\ref{fig:ENVplots}b we show the dependence on Reynolds number of the $x_*$ centroid position for both total and active regions. First, we observe that both centroids follow the same evolution. For low Reynolds numbers they are located on the left side of the sphere, at $\Rey \simeq 380$ they cross zero and for higher $\Rey$ they continue to shift upstream. This indicates that the high-shear (Couette) region near the moving wall becomes increasingly important as the Reynolds number is increased. Finally, at $\Rey=510$ almost all activity takes place within the high shear (Couette) part. This agrees with the numerical observations in plane Poiseuille flow with zero net flux that the turbulent structures move with/against the direction of the Poiseuille component for low/high Reynolds numbers (see Ref. \onlinecite{tuckerman_turbulent-laminar_2014}, note that in that paper the direction of the Poiseuille component is in the positive $x_*$ direction, opposite to our case). However, recall that instead of measuring the propagation speed of the turbulent structure, we are measuring the direction in which the turbulent spot extends. One should think of this as continuous advection of the turbulence, which decays as it moves downstream and is simultaneously continuously regenerated by a permanent perturbation.

As mentioned in the discussion of Fig.~\ref{fig:FT_Vis8a}a, for sufficiently low $\Rey$, there is no active region; the perturbations seen are vortices in the wake of the perturbing sphere, which are located downstream/left from the sphere. Since the right side ($x_*>0$) is less affected by the sphere, we plot in Fig.~\ref{fig:ENVplots}a the part of the active region located only on the right side of the sphere (green crosses). We note that the area of this portion of the active area remains nearly equal to zero up to $\Rey=330$ and then starts to grow.

The Reynolds number dependence of the streamwise and spanwise size of turbulent spots is presented in Fig.~\ref{fig:ENVplots}c,d. Both of them grow monotonically with Reynolds number up to $\Rey=470$. At $\Rey=510$ the spanwise extent seems to saturate as a result of the finite size of our test section. The streamwise extent is less affected, as the streamwise dimension of our installation is bigger ($L_x/h = 2000/5.75 = 347.8$) than the spanwise one ($L_z/h = 520/5.75 = 90.4$). The spanwise extent grows with Reynolds number by adding new streaks in the $z$ direction, which can be observed in Fig.~\ref{fig:ENV_RES}b. Similar behaviour has been observed numerically in plane Couette flow \citep{duguet_stochastic_2011}.

Finally, in order to separate the turbulent structures that move downstream and upstream, we record a video of a turbulent spot generated by the sphere for $\Rey=470$. To do this, we use the video camera with acquisition frequency $f=25$ Hz, which enables us to calculate the two dimensional spatio-temporal ($x,t$) FFT transform for each $z$ location. Motivated by the decomposition of travelling waves in thermal convection \citep{croquette_nonlinear_1989,kolodner_complex_1990}, we introduce the following procedure: we calculate the inverse FFT of the two dimensional spectrum ($k_x,\omega$) for each of the quadrants I ($k_x>0,\omega>0$) and II ($k_x<0,\omega>0$) separately. These two quadrants represent the travelling waves that go to the right/upstream (with phase $k_xx-\omega t$) and left/downstream (with phase $k_xx+\omega t$) respectively. In Fig.~\ref{fig:LeftRight} we present the resulting fields for a given instant of time. Video frames are normalized by dividing their intensity by that of the image of the laminar flow (Fig.~\ref{fig:LeftRight}a), whereas in Fig.~\ref{fig:LeftRight}b,c we plot the fluctuations of the normalized pixel intensity. Fig.~\ref{fig:LeftRight}a shows a turbulent spot with a characteristic V-shape pointing to the left. The dominant pattern of the right-going structures (Fig.~\ref{fig:LeftRight}b) are oblique waves at the tips of turbulent spot (similar to  those found in plane Poiseuille flow \citep{carlson_flow-visualization_1982,henningson_wave_1987}). In contrast, the downstream pattern (Fig.~\ref{fig:LeftRight}c) contains the wavy streaks, which defines our active region. Thus it is associated with the turbulent core of the spot (see also Supplemental Material at [URL will be inserted by publisher] for the full video showing this propagation). This difference between two patterns indicates the role of the streamwise pressure gradient (absent in plane Couette flow), which breaks the left/right symmetry. One can observe the tape joining the two ends of the plastic belt ($x_*=35$ in Fig.~\ref{fig:LeftRight}a), which at this instant moves to the left, is visible only in the pattern moving downstream/left (Fig.~\ref{fig:LeftRight}c). This confirms that our decomposition of the flow visualization separates the upstream and downstream patterns.

%Motivated by the decomposition of travelling waves in thermal convection \citep{croquette_nonlinear_1989,kolodner_complex_1990}, we introduce the following procedure: first, we calculate the Hilbert transform discarding half (quadrants III and IV) of the computed two dimensional spectrum ($k_x,\omega$). The remaining Fourier modes in quadrants I and II represent the travelling waves that go to the right/upstream (with phase $k_xx-\omega t$) and left/downstream (with phase $k_xx+\omega t$) respectively. Calculating the inverse FFT for each of these quadrants separately enables us to reconstruct the right and left going travelling waves. 

\section{Conclusions}
\label{sec:5}
 
We have presented a new installation to investigate plane Couette-Poiseuille flow. We have achieved this by combining a moving belt with a streamwise pressure gradient forcing the back-flow in the opposite direction. The mean velocity of the resulting base flow is nearly zero, which enables us to generate turbulent structures which are nearly stationary in the laboratory frame. This is the first time that stationary structures have been generated experimentally in a shear flow with a non-zero streamwise pressure gradient.

We describe the observation of the following sequence as Reynolds number is increased: laminar state, localized spots which grow with Reynolds number (both in the streamwise and spanwise directions) and finally, oblique expansion of the spot, which forms a turbulent band.

We note that we have characterized the linear stability of our particular velocity profile, given by $U(y_*)=\frac{3}{4} ({y_*}^2-1)+\frac{1}{2}(y_* +1)$ by solving the Orr-Sommerfeld/Squire equations for two dimensional, wall-bounded, parallel shear flow, using a Matlab\textregistered \, code~\citep{computer_Hoepffner} and have not found any linear instability up to $\Rey=10^8$. Our velocity profile is equivalent (under the combined operations of reflection in $x$ and $y$ and a Galilean transformation) to the profile $U(y_*)=\frac{3}{4} (1-{y_*}^2)+\frac{1}{2}(y_* +1)$, which was shown to be linearly stable \citep{balakumar_finite-amplitude_1997}, similarly to plane Couette or pipe flow.

We present the first demonstration that the transition to turbulence in plane Couette-Poiseuille flow is subcritical in nature and occurs through localized turbulent structures (spots), similarly to other shear flows such as boundary layer, pipe, pure plane Poiseuille and Couette flows. We have measured with Particle Image Velocimetry (PIV) the flow structure inside the gap, showing that the domain is divided into high-shear (Couette) and low-shear (Poiseuille) regions of activity (see Fig.~\ref{fig:STturb}a). The plane Couette-Poiseuille flow has thus only one high-shear layer near the moving wall, which differentiates it from the classical symmetric plane Poiseuille flow with two high-shear regions. We have also measured the perturbation of the flow due to the passage of a spot, which is located initially in the high-shear region near the wall. As time proceeds the spot fills the whole gap. Finally, the flow in the high-shear region becomes less turbulent but the low-shear (Poiseuille) region remains active.  In addition, we can observe (see Fig.~\ref{fig:4}a,b and Fig.~\ref{fig:STturb}c) that the spot moves to the right, which is the upstream direction with respect to the backflow induced by the pressure gradient (Poiseuille component).

We have also investigated the $\Rey$ dependence of the size of the turbulent spot triggered by a constant, localized perturbation. Two regions of the turbulent spot can be distinguished: the active turbulent core characterized by waviness of the streaks (reminiscent of travelling waves) and the total area, which also includes the weak undisturbed streaks and oblique waves at the laminar-turbulent interface. We have shown that the area of both regions and their streamwise/spanwise extents grow monotonically with Reynolds number. Analyzing the evolution of the centroid positions, we have shown that for low Reynolds numbers the turbulent spot extends downstream. As the Reynolds number is increased the spot shifts further upstream, which means that the high-shear (Couette) region becomes increasingly dominant. Similar behaviour has been observed numerically in Poiseuille flow with zero mean flux, where the turbulent structures move with/against the direction of the Poiseuille component for low/high $\Rey$ \citep{tuckerman_turbulent-laminar_2014}.

We isolate the right and left going waves, showing that the former are dominated by oblique waves located mainly at the tips of the turbulent spot, while the latter are related to the active core of the turbulent spot. This left-right symmetry breaking is due to the streamwise pressure gradient.% (absent in the case of plane Couette flow).

This new experiment to study subcritical transition to turbulence in wall-bounded flows is capable of producing high-quality detailed information on the dynamics of turbulent spots. The only existing investigation of Couette-Poiseuille flow with zero mean velocity was reported in Ref. \onlinecite{huey_plane_1974}, however they operated only in the fully turbulent regime, with sparse spatial resolution inside the gap and without visualizations. In contrast, in our installation we have measured the streamwise fluctuations (the basic flow modifications) produced during the passage of the spot by very precise PIV measurements with high spatial resolution inside the narrow gap of the facility. In addition, the wide geometry of this experimental set up gives us a large field for clear and very high contrast flow visualizations. This has allowed us to obtain the first quantitative and systematic results on the spatial evolution of turbulent spots, marking an important advance with respect to previous experiments.
%shows a very good capability to produce detailed

\begin{acknowledgements}
We thank Matthiew Chantry and Tomasz Bobinski for fruitful discussions, as well as Arnaud Prigent for help with image processing. We also acknowledge Konrad Gumowski and Tahar Amorri for technical assistance. This work was supported by a grant, TRANSFLOW, provided by the Agence Nationale de la Recherche (ANR).
\end{acknowledgements}

%\nocite{*}
%merlin.mbs apsrev4-1.bst 2010-07-25 4.21a (PWD, AO, DPC) hacked
%Control: key (0)
%Control: author (0) dotless jnrlst
%Control: editor formatted (1) identically to author
%Control: production of article title (0) allowed
%Control: page (1) range
%Control: year (0) verbatim
%Control: production of eprint (0) enabled
\providecommand{\noopsort}[1]{}\providecommand{\singleletter}[1]{#1}%
%

%\bibliography{Bibliography}% Produces the bibliography via BibTeX.
%, , aipsamp
\end{document}